\documentclass{PoS}

\usepackage{epsf}
\usepackage{amssymb}
\usepackage{amsmath}
\usepackage{amsfonts}
\usepackage{cite}
\usepackage[small]{caption2}

\usepackage[T1]{fontenc}
\usepackage{psfrag,epsfig,graphicx,graphics}

\newcommand{\be}{\begin{equation}}
\newcommand{\beq}{\begin{equation}}
\newcommand{\ee}{\end{equation}}
\newcommand{\eq}{\end{equation}}
\newcommand{\eeq}{\end{equation}}

\newcommand{\bea}{\begin{eqnarray}}
\newcommand{\eea}{\end{eqnarray}}

\def\slashchar#1{\setbox0=\hbox{$#1$}
   \dimen0=\wd0
   \setbox1=\hbox{/} \dimen1=\wd1
   \ifdim\dimen0>\dimen1
      \rlap{\hbox to \dimen0{\hfil/\hfil}}
      #1
   \else
      \rlap{\hbox to \dimen1{\hfil$#1$\hfil}}
      /gdatdafinal2.tex
   \fi}

\title{QCD factorization beyond leading twist in exclusive processes: $\rho_T$-meson production}

\ShortTitle{QCD factorization beyond leading twist in exclusive processes: $\rho_T$-meson production}

\author{I.~V.~Anikin\\
Bogoliubov Laboratory of Theoretical Physics, JINR,
             141980 Dubna, Russia\\
E-mail: \email{anikin@theor.jinr.ru}}

\author{D.~Yu.~Ivanov\\
Sobolev Institute of Mathematics, 630090 Novossibirsk, Russia\\
E-mail: \email{d-ivanov@math.nsc.ru}}

\author{B.~Pire\\
CPHT, {\'E}cole Polytechnique, CNRS, 91128 Palaiseau Cedex, France\\
E-mail: \email{pire@cpht.polytechnique.fr}}

\author{L.~Szymanowski\\
Soltan Institute for Nuclear Studies, PL-00-681 Warsaw, Poland\\
E-mail: \email{Lech.Szymanowski@fuw.edu.pl}}

\author{\speaker{S.~Wallon}
\\
        LPT, Universit{\'e} Paris-Sud, CNRS, 91405 Orsay, France \ {\em \&} \\
UPMC Univ. Paris 06, facult\'e de physique, 4 place Jussieu, 75252 Paris Cedex  05, France\\
        E-mail: \email{wallon@th.u-psud.fr}}

\abstract{Exclusive processes in hard
electroproduction are one of the best place for understanding the
factorization properties of QCD.
The HERA experiment recently provided precise data for $\rho$ electroproduction,
 including all
spin density matrix elements. From QCD, it is expected that such a process should factorize into
a hard (calculable) coefficient function, and hadronic ($p$ and $\rho$) matrix elements. Such
a factorization is up to now only proven for a longitudinaly polarized $\rho$.  
Within the $k_t$-factorization
approach (valid at large  $s_{\gamma^* p}$),
we evaluate the impact factor of the transition $\gamma^{*} \to \rho_{T}$ taking
 into account
the twist $3$ contributions. We show that
a gauge invariant expression is obtained with the help of QCD equations of motion. 
More generally, relying on these equations and 
on the invariance under rotation on the light-cone of
 the factorized amplitude, the  non-perturbative Distribution Amplitudes can be reduced to a minimal set.
 This opens the way to a
consistent treatment of factorization for exclusive processes with a transversally polarized vector meson.
We prove the equivalence of two proposed parametrizations of the $\rho_{T}$ distribution amplitudes.
}

\FullConference{European Physical Society Europhysics Conference on High Energy Physics,
EPS-HEP 2009,\\
		 July 16 - 22 2009\\
		 Krakow, Poland}

\begin{document}

\section{Introduction}
\label{Sec_Int}

Studies of hard exclusive reactions rely on the factorization properties of the leading twist amplitudes \cite{fact}.
The leading twist
distribution amplitude (DA) of a transversally polarized vector meson is
chiral-odd, and hence decouples from hard amplitudes even when another
chiral-odd quantity is involved \cite{DGP} unless in reactions with more than two final
hadrons \cite{IPST}.
Thus
transversally polarized $\rho-$meson production is generically governed by  
twist 3 contributions for which  a pure collinear
factorization fails  due to the appearance of end-point singularities \cite{MP,AT}.
The meson quark gluon structure within collinear factorization may be described by Distribution Amplitudes (DAs), 
classified in \cite{BB}. Measurements \cite{exp} of the 
$\rho_T-$meson production amplitude  in
 photo and electro-production   show that it
 is by no means negligible.
We  consider here the case of  very high energy collisions at colliders, for which
future progress  may come from real or virtual photon photon collisions \cite{IP,PSW}.
In the literature there are two approaches to the factorization of the
scattering amplitudes in exclusive processes at leading and higher twists. The first approach \cite{APT,AT}, the Light-Cone Collinear Factorization (LCCF), extends the inclusive approach \cite{EFP}
 to exclusive processes, dealing with the factorization in the momentum space around the dominant light-cone
direction.  On the other hand, there exists a Covariant Collinear Factorization (CCF) approach
in coordinate space succesfully applied in \cite{BB} for a systematic
description of DAs of hadrons carrying different
twists.  
We show  \cite{us} that these two descriptions are equivalent at twist 3, and  illustrate this by   calculating
within both methods the impact factor $\gamma^* \to \rho_T$, 
up to twist 3
accuracy. 

\section{LCCF factorization of exclusive processes}
 \label{Sec_LCCF}


The amplitude for the exclusive process $A \to \rho \, B$ is,  in
 the momentum representation and in axial
gauge reads ($H$ and $H_\mu$ are  2- and 3-parton coefficient functions,
 respectively)
\begin{eqnarray}
\label{GenAmp}
{\cal A}=
\int d^4\ell \, {\rm tr} \biggl[ H(\ell) \, \Phi (\ell) \biggr]+
\int d^4\ell_1\, d^4\ell_2\, {\rm tr}\biggl[
H_\mu(\ell_1, \ell_2) \, \Phi^{\mu} (\ell_1, \ell_2) \biggr] + \ldots \,.
\end{eqnarray}
In (\ref{GenAmp}), the soft parts $\Phi$ are  the
Fourier-transformed 2- or 3-parton correlators which are matrix elements of non-local operators.
To factorize the amplitude, we  choose the dominant direction around which
we  decompose our relevant momenta and  we Taylor expand the hard part.
Let $p\sim p_\rho$ and $n$ be two light-cone vectors ($p \cdot n =1$).  Any vector $\ell$ is then expanded as
\begin{eqnarray}
\label{k}
\ell_{i\, \mu} = y_i\,p_\mu  + (\ell_i\cdot p)\, n_\mu + \ell^\perp_{i\,\mu} ,
\quad y_i=\ell_i\cdot n ,
\end{eqnarray}
and  the integration measure in (\ref{GenAmp}) is replaced as
$d^4 \ell_i \longrightarrow d^4 \ell_i \, dy_i \, \delta(y_i-\ell\cdot n) .$
The hard part  $H(\ell)$ is then expanded around
the dominant  $p$ direction:
\begin{eqnarray}
\label{expand}
H(\ell) = H(y p) + \frac{\partial H(\ell)}{\partial \ell_\alpha} \biggl|_{\ell=y p}\biggr. \,
(\ell-y\,p)_\alpha + \ldots
\end{eqnarray}
where $(\ell-y\,p)_\alpha \approx \ell^\perp_\alpha$ up to twist 3.
To obtain a factorized amplitude, one performs 
an
 integration by parts
to replace  $\ell^\perp_\alpha$ by $\partial^\perp_\alpha$ acting on
the soft correlator.
 This leads to new operators containing
transverse derivatives, such as $\bar \psi \, \partial^\perp \psi $,
 thus requiring
additional DAs
$\Phi^\perp (l)$.
Factorization is then achieved by 
 Fierz decomposition on a set of relevant Dirac $\Gamma$ matrices, and we end up with
\bea
\label{GenAmpFac23}
\hspace{-.4cm}{\cal A}=
  {\rm tr} \left[ H_{q \bar{q}}(y) \, \Gamma \right] \otimes \Phi_{q \bar{q}}^{\Gamma} (y)
+
 {\rm tr} \left[ H^{\perp\mu}_{q \bar{q}}(y)  \Gamma \right] \otimes \Phi^{\perp\Gamma}_{{q \bar{q}}\,\mu} (y) + {\rm tr} \left[ H_{q \bar{q}g}^\mu(y_1,y_2) \, \Gamma \right] \otimes \Phi^{\Gamma}_{{q \bar{q}g}\,\mu} (y_1,y_2) \,,
\eea
%
%
where $\otimes$ is the $y$-integration.
Although the fields coordinates $z_i$ are on the light-cone in both LCCF and CCF parametrizations of the soft non-local correlators,
 $z_i$ is along $n$ in LCCF while arbitrary in CCF.
The transverse physical polarization of the $\rho-$meson is defined by the conditions
\beq
\label{pol_RhoTdef}
e_T \cdot n=e_T \cdot p=0\,.
\eq
Keeping all the terms up to the twist-$3$ order
with the axial (light-like) gauge, $n \cdot A=0$,
the matrix elements of quark-antiquark nonlocal operators
 for vector and axial-vector correlators without and with transverse derivatives,
with $\stackrel{\longleftrightarrow}
{\partial_{\rho}}=\frac{1}{2}(\stackrel{\longrightarrow}
{\partial_{\rho}}-\stackrel{\longleftarrow}{\partial_{\rho}})\,,$
can be written
as (here, $z=\lambda n$)
\begin{eqnarray}
\label{par1a}
&&\langle \rho(p_\rho)|
\bar\psi(z)\gamma_5\gamma_{\mu} \psi(0) |0\rangle =
m_\rho\,f_\rho \, i\int_{0}^{1}\, dy \,\text{exp}\left[iy\,p\cdot z\right]\varphi_A(y)\, \varepsilon_{\mu\alpha\beta\delta}\,
e^{*\alpha}_{T}p^{\beta}n^{\delta} \, \nonumber \\
\label{par1.1a}
&&\langle \rho(p_\rho)| \bar\psi(z)\gamma_5\gamma_{\mu}
i\stackrel{\longleftrightarrow}
{\partial^T_{\alpha}} \psi(0) |0\rangle =
m_\rho\,f_\rho \,
i\int_{0}^{1}\, dy \,\text{exp}\left[iy\,p\cdot z\right]\varphi_A^T (y) \, p_{\mu}\, \varepsilon_{\alpha\lambda\beta\delta}\,
e_T^{*\lambda} p^{\beta}\,n^{\delta}\,,
\end{eqnarray}
for the axial case, where 
$y$ ($\bar y$) is the quark (antiquark) momentum fraction.
Two analogous correlators are needed to describe gluonic degrees of freedom, introducing $B$ and $D$ DAs. One thus needs 7 DAs:  $\varphi_1$ (twist-$2$), 
$B$ and $D$ (genuine (dynamical) twist-$3$) and
$\varphi_3$, $\varphi_A, \varphi_1^T$, $\varphi_A^T$
(contain both parts: kinematical (\`a la
Wandzura-Wilczek) twist-$3$ and genuine (dynamical) twist-$3$).
These DAs  are related by 2 Equations of Motions (EOMs) and 2 equations arising from the invariance of  ${\cal A}$ under
rotation on the light-cone. Indeed, this invariance  with respect to $n$ does not involve the hard part of ${\cal A}$, 
and therefore implies constraints on the soft part, i.e. on the DAs. We thus have only 3 independent DAs $\varphi_1$ , 
$B$ and $D$, which fully encode the non-perturbative content of the $\rho$ at twist 3.

 The original CCF parametrizations of the $\rho$ DAs~\cite{BB} also  involve 3 independent DAs, defined through 4  correlators related by EOMs. For example, the 2-parton axial-vector correlators reads, 
\beq
\label{BBA}
\langle \rho(p_\rho)|\bar \psi(z) \, [z,\, 0] \, \gamma_\mu \gamma_5 \psi(0)|0\rangle =
\frac{1}{4}f_\rho\,m_\rho\, \varepsilon_\mu^{\,\,\,\alpha \beta \gamma} e^*_{T \alpha} \,p_\beta \, z_\gamma\,    \int\limits_0^1\,dy\,e^{iy(p \cdot z)}\,g_\perp^{(a)}(y)\;,
\eeq
%
$[z_1, \, z_2] = P \exp \left[ i g \int\limits^1_0 dt \, (z_1-z_2)_\mu A^\mu(t \,z_1 +(1-t)\,z_2    \right]$
being the Wilson line. Denoting the meson polarization vector by $e,$
 $e_T$ is here defined to be orthogonal to the light-cone vectors $p$ and $z$:
\beq
\label{pol_Rho}
e_{T \mu}=e_\mu -p_\mu \frac{e \cdot z}{p \cdot z}-z_\mu \frac{e \cdot p}{p \cdot z} \, ,
\eeq
Thus  $e_T$  (\ref{pol_Rho}) in CCF and $e_T$ (\ref{pol_RhoTdef}) in LCCF differ since  $z$ does not generally point in the $n$ direction.

\section{$\gamma^* \to \rho_T$ Impact factor up to  twist three accuracy in LCCF and CCF}

We have calculated, in both LCCF and CCF, the forward impact factor $\Phi^{\gamma^*\to\rho}$ 
of the subprocess
 $g+\gamma^*\to g+\rho_T\,,$
 defined as
the integral of the  discontinuity in the $s$
channel of the  off-shell S-matrix element 
 ${\cal S}^{\gamma^*_T\, g\to\rho_T\, g}_\mu$.
  In LCCF, one computes the diagrams perturbatively in a fairly direct way,
which makes the use of the  CCF
 parametrization \cite{BB} less practical.
 We need to express the impact factor in terms of 
hard coefficient functions and soft parts parametrized by the light-cone matrix elements. The standard technique 
here is an operator product expansion on the light cone, which  gives the leading term in the power counting.
Since there is no operator definition for an impact factor, we have to rely 
on perturbation theory.
The primary complication here is that the  $z^2\to 0$ limit of any single diagram is given in terms of  
light-cone matrix elements without any Wilson line insertion between the quark and gluon operators (''perturbative correlators``), like
$
\langle \rho(p_\rho)|\bar \psi(z)\gamma_\mu \psi(0)|0 \rangle\,.$
Despite working in the axial gauge one cannot neglect  effects coming from the Wilson
lines since the  two light cone vectors $z$ and $n$ are not identical and thus, generically, Wilson lines are not equal to unity. Nevertheless in the axial gauge the contribution of each additional parton costs one extra power 
of $1/Q$, allowing the calculation to be
organized in a simple iterative manner expanding the Wilson line. 
At twist 3, we need to keep the contribution
$[z,0]=1+i \,g \int\limits^1_0 dt \, z^\alpha A_\alpha (z t)$ and to care about the difference between  the physical  $\rho_T$-polarization (\ref{pol_RhoTdef}) from the formal one  (\ref{pol_Rho}).
At twist 3-level the net effect of the Wilson line when computing our impact factor is
just a renormalization of the DA  $g^a_\perp$  of  (\ref{BBA}), and similarly for the vector case.


Based 
 on the solution of the EOMs and $n$-independence set of equations,  our two LCCF and CCF results are identical; they are gauge invariant due to a consistent inclusion of fermionic and gluonic degrees of freedom and  are free of end-point singularities, due to the $k_T$ regulator.


 This work is partly supported by the ECO-NET program, contract
18853PJ, the French-Polish scientific agreement Polonium, the grant
ANR-06-JCJC-0084, the RFBR (grants 09-02-01149,
 08-02-00334, 08-02-00896), the grant NSh-1027.2008.2 and
the Polish Grant N202 249235.

\end{document}